\pdfoutput=1 
\documentclass[10pt,journal,compsoc]{IEEEtran}

\ifCLASSOPTIONcompsoc
  \usepackage[nocompress]{cite}
\else
  \usepackage{cite}
\fi

\usepackage{amsmath}
\usepackage{mathtools}
\usepackage{makecell}
\usepackage{bm}
\usepackage{multirow}
\usepackage{makecell}

\usepackage{color}

\newcommand{\mytilde}{\raise.17ex\hbox{$\scriptstyle\sim$}}
\usepackage{threeparttable}

\begin{document}
\bstctlcite{IEEEexample:BSTcontrol}

\title{Leveraging Highly Approximated Multipliers\\ in DNN Inference}

\author{
Georgios~Zervakis,
Fabio Frustaci,~\IEEEmembership{Member,~IEEE,}
Ourania~Spantidi,\\
Iraklis~Anagnostopoulos,~\IEEEmembership{Member,~IEEE,}
Hussam~Amrouch, \IEEEmembership{Member, IEEE},
J\"org~Henkel, \IEEEmembership{Fellow, IEEE}%

\IEEEcompsocitemizethanks{\IEEEcompsocthanksitem G. Zervakis is with the Computer Engineering \& Informatics Dept., University of Patras, Greece.
Email: zervakis@ceid.upatras.gr
\IEEEcompsocthanksitem F.~Frustaci is with the DIMES, University of Calabria, Italy.
Email: f.frustaci@dimes.unical.it
\IEEEcompsocthanksitem O.~Spantidi is with the Department of Computer Science at Eastern Michigan University, USA.
Email: ourania.spantidi@emich.edu
\IEEEcompsocthanksitem I.~Anagnostopoulos is with the School of Electrical, Computer and Biomedical Engineering, Southern Illinois University Carbondale, USA.
Email: iraklis.anagno@siu.edu
\IEEEcompsocthanksitem H.~Amrouch is with the Chair of AI Processor Design, TUM School of Computation, Information and Technology and Munich Institute of Robotics and Machine Intelligence, Technical University of Munich (TUM), Germany.
Email: amrouch@tum.de
\IEEEcompsocthanksitem J. Henkel is with the Chair for Embedded Systems, Karlsruhe Institute of Technology, Germany.
Email: henkel@kit.edu}%
\thanks{Corresponding author: Georgios Zervakis (zervakis@ceid.upatras.gr).}%
}

\IEEEtitleabstractindextext{%
\begin{abstract}
In this work, we present a control variate approximation technique that enables the exploitation of highly approximate multipliers in Deep Neural Network (DNN) accelerators.
Our approach does not require retraining and significantly decreases the induced error due to approximate multiplications, improving the overall inference accuracy.
As a result, our approach enables satisfying tight accuracy loss constraints while boosting the power savings.
Our experimental evaluation, across six different DNNs and several approximate multipliers, demonstrates the versatility of our approach and shows that compared to the accurate design, our control variate approximation achieves the same performance, 45\% power reduction, and less than 1\% average accuracy loss.
Compared to the corresponding approximate designs without using our technique, our approach improves the accuracy by 1.9x on average.
\end{abstract}
}

\maketitle
\IEEEpeerreviewmaketitle

\IEEEraisesectionheading{\section{Introduction}}

\IEEEPARstart{D}{eep} Neural Networks (DNNs) have become one of the main methodologies to enable artificial intelligence (AI) in many applications fields~\cite{jouppi2017datacenter}.
The promising results offered by DNNs derive from the processing of a huge amount of data that, in many cases where a real time response is crucial, prevents their software-based execution~\cite{jouppi2017datacenter}.
Customized hardware DNN accelerators meet the demand for high inference speed.
This is particularly emphasized when the DNN is deployed into IoT edge devices, where the computation has to be performed locally with a reduced resources budget.
Multiply-accumulate (MAC) operation is the most intensive computational task performed by a DNN during the inference phase.
As an example, the most popular Convolutional Neural Networks (CNNs) perform millions of MAC operations in their convolutional and fully-connected layers.
In order to speed-up the inference phase, DNN accelerators are typically equipped with thousands of MAC units (e.g., $4$K MAC units in the Google Edge TPU)  operating in parallel, thus leading to high power requirements~\cite{amrouch2020npu,Zervakis:TC2022}.
Since the MAC operations are responsible for most of the power consumption, the research effort has focused on optimizing the multiplication, which is the most complex operation in a MAC.

Recently, approximate computing has emerged as a powerful design paradigm that relaxes the constraint of an exact computation in error-resilient applications, in order to trade the quality of the result with speed, area, and power consumption~\cite{Alioto_approximate,ZervakisTVLSI2019}.
Due to their inherent error-tolerance, DNNs have become appropriate candidates for approximate computations~\cite{sarwar2018energy,mrazek2016design,tasoulas2020weight,alwann,zervakis2020design,spantidi2021pene, Spantidi:TETC2023, Zervakis:TC2022}.
Voltage scaling can help reduce power consumption and manage temperature; however, it significantly decreases the throughput of DNNs~\cite{amrouch2020npu, song20197} while voltage over-scaling~\cite{Paim:TCASI2021} can lead to unpredictable timing errors, which may severely impact accuracy~\cite{Salamin:DATE2021, Zervakis:TVLSI2018}.

Approximating a DNN imposes several challenges.
The layers within the same DNN can have a significantly different error-resiliency~\cite{tasoulas2020weight,alwann,zervakis2020design,spantidi2021pene}.
Moreover, the errors due to approximate circuits are not constant but they are highly input dependent~\cite{zervakis2020design}.
Finally, it has been shown that the deeper the neural network, the more sensitive it becomes to even slight approximation~\cite{tasoulas2020weight}.
State-of-the-art applies retraining to mitigate the accuracy loss due to approximation~\cite{sarwar2018energy,mrazek2016design}.
However, retraining may be infeasible in many cases because it is either time consuming or the training set might not be available (e.g., proprietary models)~\cite{alwann}.
Many approximate multipliers have been proposed in the past, such as those based on column truncation~\cite{Solaz_TCAS2012,FrustaciTCASII2020,Mrazek_TVLSI_2018}, approximate compressors~\cite{Pei_TCASII_21_apprCompress,Strollo_approxCompres}, partial product perforation~\cite{ZervakisTVLSI2016}, and recursive multipliers with approximate blocks~\cite{kulkarni_appr_blocks,WarisTETC2021:axrm}.
Nevertheless, employing such multipliers in DNN inference, without retraining, results to unacceptable accuracy degradation~\cite{spantidi2021pene}, even when the slightest approximation is selected.

To enable effective exploitation of approximate multipliers in DNN inference and maximize power reduction, we propose a \textit{control variate approximation} method~\cite{zervakis2021control}.
Our technique improves the accuracy of approximate DNN accelerators by estimating and mitigating at run-time the error caused by approximate multiplications, without the need to perform time-overwhelming retraining.
Leveraging the accuracy improvement of our control variate approximation, we are able to integrate highly approximate multipliers in DNN inference, and thus maximize the achieved power reduction.
Our extensive experimentation demonstrates that our technique improves the inference accuracy by $1.9$x on average, when compared to exactly the same approximate DNN accelerator without our proposed control variate approximation.
This is an expanded version of our work~\cite{zervakis2021control}.
\textit{In~\cite{zervakis2021control} control variate was bound to only a specific approximate multiplier~\cite{ZervakisTVLSI2016}.}
In this work, we extend~\cite{zervakis2021control} and we demonstrate how our control variate approximation can be employed with a diversity of approximate multipliers by adjusting accordingly our rigorous mathematical formulation of the induced convolution error. 
Overall, we demonstrate the efficiency and versatility of our technique when considering the approximate perforated, recursive, and truncated multipliers.
Such multipliers, apply aggressive approximation by inducing high error, but also achieving high power reduction.
We designed approximate DNN accelerators based on the developed control variate equations and we performed a complete comparison among the investigated approximate multipliers, demonstrating that our control variate approximation is able to mitigate most of the accuracy loss while maintaining high power gains.

\section{Description of Approximate Multipliers}\label{sec:axmults}

This section provides a brief description of three approximate multipliers that will be used in our work to generate approximate DNN accelerators and evaluate the efficiency of our control variate method in mitigating the error due to the approximate multiplications.
In our analysis we consider partial product perforated multipliers~\cite{ZervakisTVLSI2016}, truncated multipliers~\cite{Solaz_TCAS2012,FrustaciTCASII2020,Mrazek_TVLSI_2018}, and approximate recursive multipliers~\cite{kulkarni_appr_blocks,WarisTETC2021:axrm}.
The examined circuits are representative paradigms of approximate multipliers and achieve very high power reduction, at the cost, however, of high error.

\subsection{Approximate Perforated Multipliers}
The partial product perforation technique is based on the following principle.
Let us consider the multiplication operation between two $n$-bit inputs, $W$ and $A$.
The accurate result $W\cdot A$ is the sum of all the partial products, obtained by multiplying $W$ by each bit $a_i$ of $A$, with $a_i$=0...$n-1$:
\begin{equation}\label{eq:product}
W\cdot A=\sum_{i=0}^{n-1}{W\cdot a_i\cdot 2^i},
\end{equation}

The partial product perforation technique approximates the multiplication operation by omitting $m$ consecutive partial products starting from the $s$-th one, with $s\!<\!n$ and $m<n-s$.
Hence, the perforated product equals:
\begin{equation}\label{eq:axm_perf}
\mathrm{AM_P}(W,A)=\sum_{\substack{i=0, \\
i\not\in[s, s+m)}}^{n-1}{W\cdot a_i\cdot 2^i}.
\end{equation}
The authors in~\cite{ZervakisTVLSI2016} deduced that when the distribution of one of the multiplicands is unknown, $s=0$ should be preferred.
Hence, in our work we set $s=0$ and we examine varying values for $m$.
The error of the Perforated Approximate Multiplication, when the $m$ least partial products are perforated, is therefore calculated as follows:
\begin{equation}\label{eq:err_perf}
\begin{split}
\epsilon &= W\cdot A - \mathrm{AM_P}(W,A) \\
		 &= W\cdot A - W\cdot (A - A\text{ mod }2^m) \\
		 &= W\cdot p,\quad p = A\text{ mod }2^m
\end{split}
\end{equation} 

\begin{figure}[t!]
\centering
\resizebox{0.90\columnwidth}{!}{\includegraphics{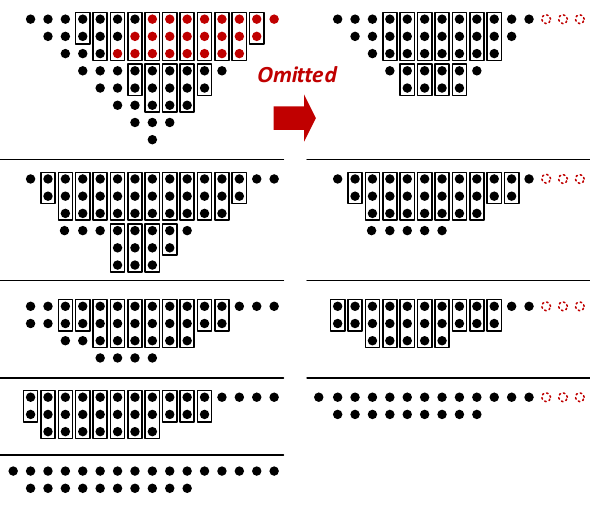}}
\caption{Partial product reduction stages for: a) the accurate multiplier b) the approximate perforated multiplier with $m=3$ and $s=0$.}
\label{fig:perforated}
\end{figure}

As an example, Fig.~\ref{fig:perforated} depicts the partial product reduction process of an unsigned $8\times 8$ multiplier without (left) and with (right) perforation.
For the perforated multiplier $m=3$ (and $s=0$) is considered.
As shown, in Fig.~\ref{fig:perforated}, compared to the exact counterpart (i.e., without perforation), the perforated multiplier has shallower partial product reduction stages and, consequently, a lower number of compressors entailing lower energy consumption and potentially higher speed. 

\subsection{Approximate Recursive Multipliers}
\begin{figure}[t!]
\centering
\resizebox{0.90\columnwidth}{!}{\includegraphics{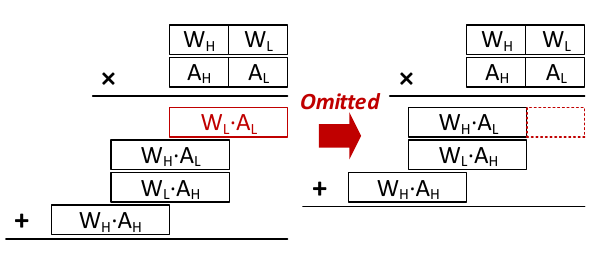}}
\caption{The principle of the Recursive Approximate Multiplier: composing a large multiplier by using smaller inaccurate building blocks}
\label{fig:low-part-block}
\end{figure}

The principle of the Recursive Approximate Multiplier is illustrated in Fig.~\ref{fig:low-part-block} (left).
Overall, a recursive multiplier decomposes the large multiplication to smaller products and accumulates them to obtain the final result.
Such a process is possibly iterated by dividing the obtained blocks into smaller ones.
One way to trade the energy consumption with the accuracy of the result is to simplify the design of the smaller blocks, employed to calculate the least significant bits of the result, by approximating their logic function. In that way,  the number of employed logic gates is reduced.
As an example,~\cite{kulkarni_appr_blocks} proposes to use an approximate $2\times 2$ multiplying block where only the multiplication between the binary inputs ``11'' and ``11'' is approximated to the inaccurate result ``111''.
Nevertheless, several partitionings can be employed for the decomposition~\cite{WarisTETC2021:axrm}.

In our work, we consider a recursive multiplier in which each input is divided into two sub-words: the low-part, composed of $m$ bits, and the high-part, composed of $n-m$ bits,  with $m<n$.
In this way, the $n\times n$ multiplier is obtained by accumulating the four sub-products as follows:
\begin{equation}
    W\cdot A = W_H \!\cdot\! A_H \!\cdot\! 2^{2m} \!+\! (W_H \!\cdot\! A_L \!+\! W_L \!\cdot\! A_H) \!\cdot\! 2^{m} \!+\! W_L \!\cdot\! A_L.
\end{equation}
Applying coarse approximation, we generate approximate recursive multipliers by pruning the entire sub-product of the two low-parts (Fig.~\ref{fig:low-part-block} (right)).
Hence, the approximate product is given by:
\begin{equation}\label{eq:axm_rec}
\mathrm{AM_R}(W,A)=(W_H \cdot A_H \cdot 2^m + W_H \cdot A_L + W_L \cdot A_H) \cdot 2^{m}
\end{equation}
and thus, the multiplication error when the size of the low-part is $m$ bits, can be calculated as follows:
\begin{equation}\label{eq:err_rec}
\begin{split}
\epsilon &= W\cdot A - \mathrm{AM_R}(W,A) \\
         &= (W_H \!\cdot\! A_H \!\cdot\! 2^m \!+\! W_H \!\cdot\! A_L \!+\! W_L \!\cdot\! A_H) \!\cdot\! 2^{m} \!+\! W_L \!\cdot\! A_L \\
         &\qquad- (W_H \cdot A_H \cdot 2^m + W_H \cdot A_L + W_L \cdot A_H) \cdot 2^{m} \\
		 &= W_L \cdot A_L
\end{split}
\end{equation} 

\subsection{Approximate Truncated Multipliers}
\begin{figure}[t!]
\centering
\resizebox{0.90\columnwidth}{!}{\includegraphics{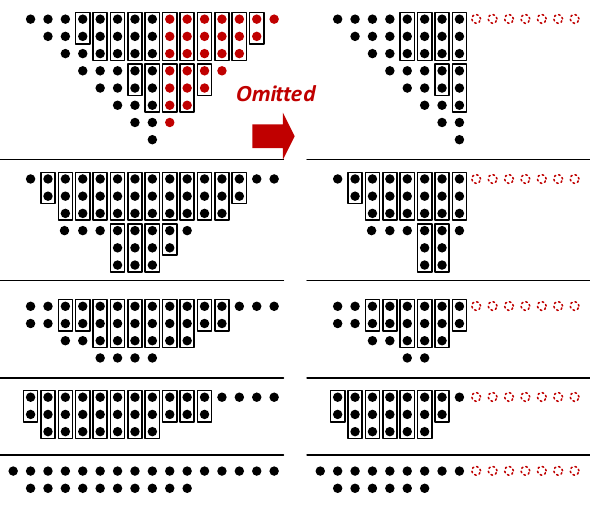}}
\caption{The truncated multiplier with $m = 7$}
\label{fig:truncation}
\end{figure}

One of the most widely used approximation techniques is truncation~\cite{Solaz_TCAS2012,FrustaciTCASII2020,Mrazek_TVLSI_2018}.
The Truncated Approximate Multiplier is composed by pruning the hardware resources that form the $m$ least significant columns of the multiplier.
At the hardware level, the \texttt{AND} gates computing the bits $w_i \cdot a_j$, with $i,j \in [0, m)$ and $i+j<m$, are not implemented in the partial product generation stage.
Consequently, all the compressors belonging to the least-significant $m$ columns of the partial product reduction stages are removed.
The application of the truncation technique with $m=7$ on an unsigned $8\times8$ multiplier is shown in Fig.~\ref{fig:truncation}.
The reduced number of partial product bits entails a lower number of compressors w.r.t. the accurate multiplier, thus leading to a lower energy consumption and potentially lower delay.
The approximate result of the Truncated Approximate Multiplier with $m$ truncated columns can be obtained by:  
\begin{equation}\label{eq:axm_trunc}
\mathrm{AM_T}(W,A)=\sum_{i=0}^{n-1}\sum_{j=\text{max}(m-i,0)}^{n-1}{w_j\cdot a_i\cdot 2^{(i+j)}}
\end{equation} 
whereas its error is computed by:
\begin{equation}\label{eq:err_trunc}
\begin{split}
\epsilon &= W\cdot A - \mathrm{AM_T}(W,A) \\
		 &= \sum_{i=0}^{n-1}\sum_{j=0}^{n-1}{w_j\cdot a_i\cdot 2^{(i+j)}} \\
		 &\qquad- \sum_{i=0}^{n-1}\sum_{j=\text{max}(m-i,0)}^{n-1}{w_j\cdot a_i\cdot 2^{(i+j)}}\\
		 &= \sum_{i=0}^{m-1}{(W \text{ mod }2^{m-i})\cdot a_i\cdot 2^i} 
\end{split}
\end{equation} 

\begin{table}[t!]
\renewcommand{\arraystretch}{1}
\caption{Error analysis on the examined approximate multipliers}
\label{tab:axmult_analysis}
  \footnotesize
\centering
  \begin{tabular}[t]{c|c|c||c|c|c}
 \hline
 \multicolumn{6}{c}{\textbf{Approximate Perforated Multiplier}} \\ \hline
\multicolumn{3}{c||}{\textbf{Unif. Distr.} $U(0,255)$} & \multicolumn{3}{c}{\textbf{Norm. Dist. $\mathcal{N}(125,24^2)$} } \\ \hline
$m$ & $\mu$ & $\sigma$ & $m$ & $\mu$ & $\sigma$\\ \hline
1 & 63.7  & 82     & 1   & 62.4    & 64.7\\ \hline
2 & 191   & 198    & 2   & 187   & 146\\ \hline
3 & 447   & 425    & 3   & 435   & 302 \\ \hline 
  \multicolumn{6}{c}{\textbf{Approximate Recursive Multiplier}} \\ \hline
\multicolumn{3}{c||}{\textbf{Unif. Distr.} $U(0,255)$} & \multicolumn{3}{c}{\textbf{Norm. Dist. $\mathcal{N}(125,24^2)$} } \\ \hline
$m$ & $\mu$ & $\sigma$ & $m$ & $\mu$ & $\sigma$\\ \hline
2 & 2.24  & 2.67     & 2   & 2.25  & 2.68 \\ \hline
3 & 12.26 & 12.51    & 3   & 12.24 & 12.47\\ \hline
4 & 56  & 53.4     & 4   & 56.2    & 53.4 \\ \hline
5 & 239   & 219      & 5   & 239   & 219\\ \hline
 \multicolumn{6}{c}{\textbf{Approximate Truncated Multiplier}} \\ \hline
\multicolumn{3}{c||}{\textbf{Unif. Distr.} $U(0,255)$} & \multicolumn{3}{c}{\textbf{Norm. Dist. $\mathcal{N}(125,24^2)$} } \\ \hline
$m$ & $\mu$ & $\sigma$ & $m$ & $\mu$ & $\sigma$\\ \hline
4 & 12   & 9.9   & 4   & 12.6 & 9.9\\ \hline
5 & 32   & 23    & 5   & 32.2   & 23 \\ \hline
6 & 80   & 52    & 6   & 80.6   & 52.8\\ \hline
7 & 192  & 115   & 7   & 192  & 127\\ \hline
  \end{tabular}
\end{table}

\subsection{Error Analysis}
The three typologies of approximate multipliers, analyzed above, have their own peculiarities in terms of accuracy.
In Table~\ref{tab:axmult_analysis}, we assess the error characteristics of the examined approximate multipliers considering varying approximation levels (i.e., $m$ values) and $8$-bit unsigned operands with uniform and normal distribution.
For the normal distribution, we arbitrarily chose, without loss of generality, a mean value and a standard variation of $125$ and $24$, respectively.
Table~\ref{tab:axmult_analysis} collects the error results obtained for 1M input operands couples.
Note that for all the approximate multipliers, as $m$ increases the applied approximation increases.
However, there is not a direct comparison between the $m$ value of different approximate multipliers.
As shown in Table~\ref{tab:axmult_analysis}, the three multipliers behave differently as their configuration knob ($m$) varies.
For example, the Approximate Perforated multiplier shows the highest error mean value ($\mu$) and highest standard deviation ($\sigma$) for both input distributions.
The $\mu$ and $\sigma$ of the Truncated Approximate Multiplier are low and scale more gracefully compared to the other multipliers.
Moreover, the latter multiplier also exhibits the lowest value of the $\sigma$/$\mu$ ratio, thus resulting in being the multiplier with the lowest coefficient of variation.
Hence, the truncated multipliers feature the lowest error dispersion overall.
Interestingly, the error performance of the truncated and the recursive approximate multipliers does not show a sensible variation as the distribution of the inputs changes.
Indeed, the values of the error $\mu$ and $\sigma$ are practically the same for both distributions in Table~\ref{tab:axmult_analysis}.

\section{Control Variate Approximation} \label{sec:error}

This section presents our control variate approximation technique and describes how it is applied with different approximate multipliers.
The approximate multipliers of Section~\ref{sec:axmults} are considered and our rigorous error analysis demonstrates how the control variate parameters can be tuned with respect to the multiplier used to minimize the error at convolution level.

The core operation of a convolution is given by:
\begin{equation}\label{eq:conv}
G=B+\sum_{j=1}^{k}{W_j\cdot A_j},
\end{equation}
where $B$ is the bias of the neuron, $W_j$ are the weights, and $A_j$ are the input activations.

Aiming for low-power operation, we replace all the accurate multipliers of the DNN accelerator with approximate ones.
We denote $\epsilon_j$ the multiplication error of the product $W_j\cdot A_j$.
Hence, $\epsilon_j$ equals the difference between the accurate and the approximate products:
\begin{equation}\label{eq:err_axmult}
\epsilon_j = W_j\cdot A_j - \mathrm{AM}(W_j,A_j).
\end{equation}
For example, assuming the approximate multipliers described in Section~\ref{sec:axmults}, 
$\epsilon_j$ can be computed using ~\eqref{eq:axm_perf},~\eqref{eq:axm_rec}, or~\eqref{eq:axm_trunc}.

Given~\eqref{eq:conv} and~\eqref{eq:err_axmult}, the convolution error $\epsilon_G$ equals:
\begin{equation}\label{eq:converror}
\begin{split}
\epsilon_G &= B+\sum_{j=1}^{k}{W_j\cdot A_j}-B-\sum_{j=1}^{k}{\mathrm{AM}(W_j,A_j)} \\
           &= \sum_{j=1}^{k}{\epsilon_j}.
\end{split}
\end{equation}

The error value of an approximate multiplier can be considered as a random variable, and is therefore defined by its mean value and variance~\cite{LiVAR3}. Denoting by $\mu_{AM}$ and $\sigma^2_{AM}$ the mean error and the error variance of the approximate multiplier $\mathrm{AM}$, the mean and variance of the approximate convolution error operation are given by:
\begin{equation}\label{eq:meanvar}
\begin{split}
\mathrm{E}[\epsilon_G] &=\mathrm{E}\big[\sum_{j=1}^{k}{\epsilon_j}\big]=k\mu_{AM}\\
\mathrm{Var}(\epsilon_G) &=\mathrm{Var}\Big(\sum_{j=1}^{k}{\epsilon_j}\Big)=k\sigma^2_{AM}.
\end{split}
\end{equation}
Note that, the error values $\epsilon_j$ are independent variables and thus their covariance is zero~\cite{LiVAR3,tasoulas2020weight}.

Hence, even if the approximate multiplier features small error (small $\mu_{AM}$ and $\sigma^2_{AM}$), the convolution error is significantly higher since it is proportional to the filter's size as \eqref{eq:meanvar} demonstrates.
In~\cite{tasoulas2020weight}, approximate multipliers with systematic error are employed and a constant correction term is used to compensate for the mean error (i.e., $\mathrm{E}[\epsilon_G]$).
However, even in this case, the error of the convolution is still high, since it is defined by its high variance ($Var(\epsilon_G)$).

In our work, we propose the utilization of a control variate technique to reduce the convolution error.
A control variate is an easily evaluated random variable, with known mean, that is highly correlated with our variable of interest.
To implement our control variate approximation technique, to perform the convolution  we compute~\eqref{eq:axconv} instead of~\eqref{eq:conv}.
\begin{equation}\label{eq:axconv}
G^\ast=B+\sum_{j=1}^{k}{\mathrm{AM}(W_j,A_j)}+V,
\end{equation}
where $V$ is the control variate.

Inspired by the convolution error $\epsilon_G$ in~\eqref{eq:converror} and targeting low computational complexity, we express the control variate $V$ as a first order polynomial: 
\begin{equation}\label{eq:v}
\begin{split}
V &= \sum_{j=1}^{k}{v_j}+C_0 \\
  &= \sum_{j=1}^{k}{x_j\cdot C_j}+C_0,\quad v_j=x_j\cdot C_j.
\end{split}
\end{equation}
where $C_j$, $\forall j\geq0$, are constants and $x_j$ is an input-dependent variable (i.e., obtained at runtime).
Obviously, setting $v_j = \epsilon_j$ and $C_0=0$ would deliver accurate results.
Nevertheless, this would neglect any hardware gains originating by the approximate multiplications due to the high computational complexity of precisely computing $e_j$ at runtime.
For example, assume that the perforated approximate multiplier~\cite{ZervakisTVLSI2016} is used.
Considering~\eqref{eq:err_perf}, if we set $C_j=W_j$, $\forall j>0$, $C_0=0$, and $x_j = p_j$, then $v_j = \epsilon_j$ and thus $\epsilon_G=V$.
Hence, in~\eqref{eq:axconv} the error of the approximate multiplications is cancelled out by the control variate $V$, leading to accurate computation of the convolution operation.
However, calculating $p_j\cdot W_j$ is computationally expensive and neglects the gains (area, power) of the perforated multiplier, since calculating $V$ requires $k$ multiplications and $k-1$ additions, and computing $p_j\cdot W_j$ requires the generation and addition of the partial products that were initially omitted. 

However, since a control variate must be easily evaluated, we simplify $V$ in~\eqref{eq:v} by setting $C_j=C$, $\forall j>1$.
As a result, $V$ is given by:
\begin{equation}\label{eq:v2}
V=C\cdot\sum_{j=1}^{k}{x_j} + C_0\text{ and }  v_j=x_j\cdot C.
\end{equation}
Note that to calculate $V$ in~\eqref{eq:v2}, only $k-1$ additions and $1$ multiplication are required.

Given~\eqref{eq:v2}, the approximate convolution using our control variate method~\eqref{eq:axconv} is written as:
\begin{equation}\label{eq:axconv2}
\begin{split}
G^\ast &=B+\sum_{j=1}^{k}{\Big(W_j\cdot A_j-\epsilon_j\Big)} + \sum_{j=1}^{k}{v_j} + C_0\\
       &=G - \sum_{j=1}^{k}{\Big(\epsilon_j - v_j \Big)}  + C_0
\end{split}
\end{equation}
and thus, the approximate convolution error equals:
\begin{equation}\label{eq:axconverr}
\epsilon_{G^\ast}= G-G^\ast=\sum_{j=1}^{k}{\Big(\epsilon_j - v_j \Big)} - C_0.
\end{equation} 

\subsection{Control Variate with Approximate Perforated Multipliers}\label{subsec:covarperf}
First, we examine the perforated multipliers~\cite{ZervakisTVLSI2016} to perform the approximate multiplication, i.e., $\mathrm{AM}(W_j,A_j) = \mathrm{AM_P}(W_j,A_j)$.
Considering the error of an approximate perforated multiplication that is given by~\eqref{eq:err_perf}, we set $x_j$ in our control variate~\eqref{eq:v2} equal to:
\begin{equation}\label{eq:v_perf}
x_j = A_j\text{ mod }2^m = A_j\, \&\, (2^m-1)
\end{equation}
and thus,~\eqref{eq:axconverr} becomes:
\begin{equation}\label{eq:axconverr_perf}
\begin{split}
\epsilon_{G^\ast}&=\sum_{j=1}^{k}{\Big(\epsilon_j - v_j \Big)} - C_0 \\
                 &=\sum_{j=1}^{k}{\Big(x_j\cdot(W_j-C) \Big)} - C_0.
\end{split}            
\end{equation} 

\begin{figure}[t!]
\centering
\resizebox{0.85\columnwidth}{!}{\includegraphics{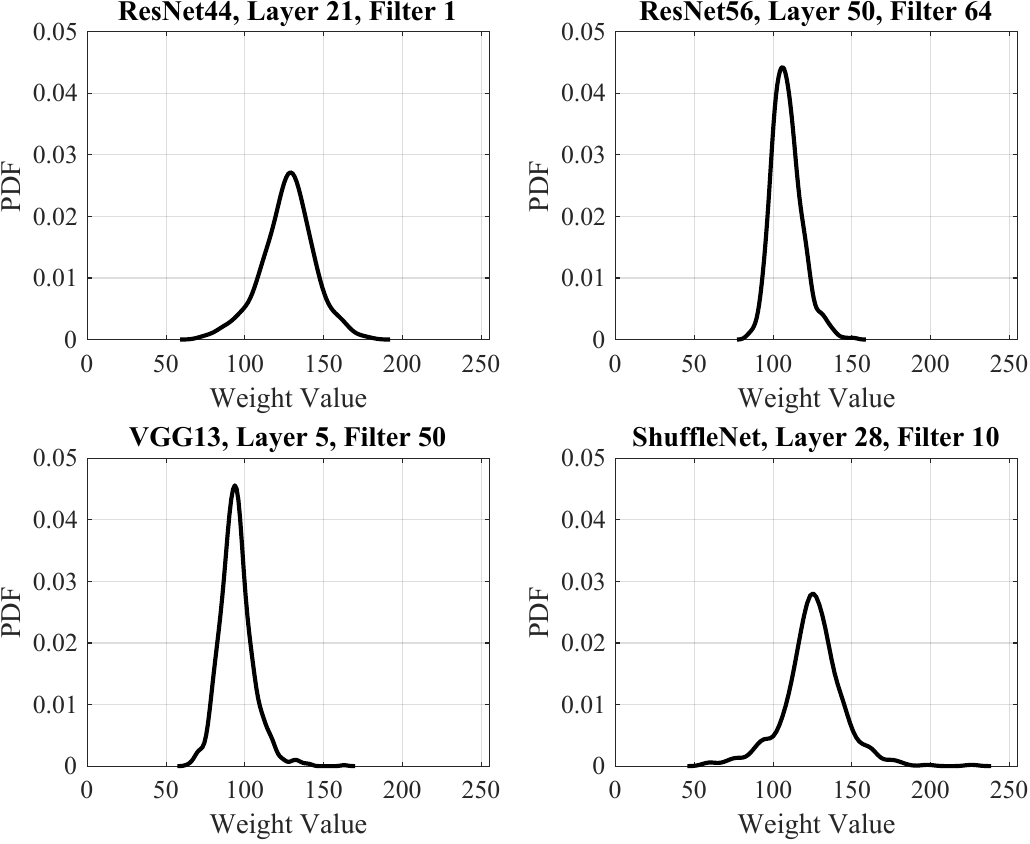}}
\caption{Weight distribution of randomly selected filters of various NNs. Four examples are depicted.
Figure obtained from~\cite{zervakis2021control}.}
\label{fig:Wdistr}
\end{figure}

Therefore, the variance $\mathrm{Var}(\epsilon_{G^\ast})$ of the error of the approximate convolution, i.e., $\epsilon_{G^\ast}$, is calculated as:
\begin{equation}\label{eq:var_perf}
\begin{split}
\mathrm{Var}(\epsilon_{G^\ast}) & = \sum_{j=1}^{k}{\mathrm{Var}\Big(\epsilon_j - v_j \Big)} \\
                       & = \sum_{j=1}^{k}{\Big((W_j-C)^2\cdot \mathrm{Var}(x_j)\Big)} \\
                       & = \underbrace{\frac{(2^m-1)(2^m+1)}{12}}_{\mathrm{Var}(x_j)}\sum_{j=1}^{k}{(W_j-C)^2}.
\end{split}
\end{equation} 
As a result, $\mathrm{Var}(\epsilon_{G^\ast})$ is minimized when:
\begin{equation}\label{eq:minvar_perf}
\begin{gathered}
\frac{d}{dC}\mathrm{Var}(\epsilon_{G^\ast}) = 0 \Rightarrow 
C=\mathrm{E}[W_j]=\frac{1}{k}\sum_{j=1}^{k}{W_j}.
\end{gathered}
\end{equation} 
Note that $C\neq 0$, i.e., variance without our control variate (as in~\eqref{eq:meanvar}).
In addition, note that the more squeezed the distribution of the weights is (i.e., concentrated close to $\mathrm{E}[W_j]$) the closer $\mathrm{Var}(\epsilon_{G^\ast})$ is to zero.
Fig.~\ref{fig:Wdistr}, shows the distribution of weights for four different examples.
In Fig.~\ref{fig:Wdistr}, the neural networks and the respective filters and layers, were randomly selected out of the neural networks we consider in Section~\ref{sec:experimental}.
Similar results are obtained for the rest of the filters and neural networks.
As shown in Fig.~\ref{fig:Wdistr}, for all the examined filters, the majority of the weights is well concentrated in a closed region (squeezed dispersion in Fig.~\ref{fig:Wdistr}).
Hence, this feature boosts the efficiency of our variance reduction method, as explained above.

Using the $C$ value obtained in~\eqref{eq:minvar_perf}, that minimizes the variance, we compute the mean convolution error $\mathrm{E}[\epsilon_{G^\ast}]$:
\begin{equation}\label{eq:axconvmean_perf}
\begin{split}
\mathrm{E}[\epsilon_{G^\ast}] &= \sum_{j=1}^{k}{\mathrm{E}\Big[\epsilon_j - v_j \Big]} - C_0\\
                     &= \sum_{j=1}^{k}{\mathrm{E}[x_j]\cdot W_j} - \sum_{j=1}^{k}{\mathrm{E}[x_j]\cdot \mathrm{E}[W_j]} - C_0\\
                     &= \underbrace{\frac{(2^m-1)}{2}}_{\mathrm{E}[x_j]}\Big(\sum_{j=1}^{k}{W_j}- k \cdot \mathrm{E}[W_j]\Big) - C_0\\
                     &= -C_0
\end{split}
\end{equation}
Therefore, by setting $C_0=0$,~\eqref{eq:axconvmean_perf} becomes zero.
As a result, the proposed control variate approximation method with $V=\mathrm{E}[W_j]\sum_{j=1}^{k}{x_j}$, effectively nullifies the mean error of the approximate convolution and also manages to decrease its variance.
Hence, the error distribution is constrained in a squeezed region around zero and high convolution accuracy is expected.
However, as~\eqref{eq:var_perf} shows, the larger $m$ is, the larger the error variance will be and thus the accuracy loss.

\subsection{Control Variate with Approximate Truncated Multipliers}\label{subsec:covartrunc}
Next, we examine the application of our proposed control variate approximation when the truncated multiplier is used for the approximate multiplication, i.e., $\mathrm{AM}(W_j,A_j) = \mathrm{AM_T}(W_j,A_j)$.
Although the error of a perforated multiplication can be easily obtained simply by the $n \times m$ product, this is not the case for the truncated multipliers.
As~\eqref{eq:axm_trunc} shows, precise error estimation of a truncated multiplication requires $m$ multiplications ($m \times 1$ down to $1 \times 1$) and $m-1$ additions. 
Hence, precise error estimation at run-time is very computationally expensive.
On the other hand, as discussed in Section~\ref{sec:axmults}, unlike the perforated multipliers, the truncated multipliers feature low error variance since it is constrained by the number of truncated columns (i.e., $m$).
Leveraging the low error variance, we can efficiently estimate the truncated multiplication error by its average value.
Considering~\eqref{eq:err_trunc}, the average error of the truncated multiplication $\mathrm{AM_T}(W_j,A_j)$, $\forall A_j$ is calculated by:
\begin{equation}\label{eq:avger_tr}
\begin{split}
\mathrm{E}[\mathrm{AM_T}(W_j,A_j)] &= \sum_{i=0}^{m-1}{\mathrm{E}[(W_j \text{ mod }2^{m-i})\cdot a_i\cdot 2^i]}\\
 & =\frac{1}{2}\sum_{i=0}^{m-1}{(W_j \text{ mod }2^{m-i})\cdot 2^i}.
\end{split}
\end{equation}

Hence, by denoting
\begin{equation}
\widehat{W}_j=\frac{1}{2}\sum_{i=0}^{m-1}{(W_j \text{ mod }2^{m-i})\cdot 2^i},
\end{equation}
the error $e_j$ of the truncated multiplication $\mathrm{AM_T}(W_j,A_j)$ is estimated by:
\begin{equation}\label{eq:e_tr}
\begin{gathered}
e_j\approx\tilde{e}_j= x_j \cdot \widehat{W}_j  \\
\text{with } x_j = (1-\delta_{0,y_j}),\quad y_j = A_j\text{ mod }2^m
\end{gathered}
\end{equation}
where $\delta_{0,y_j}$ is the Kronecker delta and thus $x_j$ is easily calculated by a logic OR of the $m$ LSB of $ A_j$.
In other words, if a multiplication error occurs, $x_j$ is $1$ while $x_j$ is $0$ when the error of the truncated multiplication is zero.
The former results to $\tilde{e}_j=\widehat{W}_j$ and the latter to $\tilde{e}_j=0$.

Given $\tilde{e}_j$ in~\eqref{eq:e_tr}, we set $V$ similarly to Section~\ref{subsec:covarperf}:
\begin{equation}\label{eq:v_tr}
\begin{gathered}
V=C\cdot\sum_{j=1}^{k}{x_j} + C_0,\quad  v_j=x_j\cdot C,\\
x_j = (1-\delta_{0,y_j}) \text{ with } y_j = A_j\text{ mod }2^m, \\
C = \mathrm{E}[\widehat{W}_j] = \frac{1}{k}\sum_{j=1}^{k}{\widehat{W}_j}.
\end{gathered}
\end{equation}

Thus, in the case of the approximate truncated multiplier, the convolution error~\eqref{eq:axconverr} becomes:
\begin{gather}
\begin{aligned}
\epsilon_{G^\ast}\!&=\!\sum_{j=1}^{k}{\Big(\epsilon_j - v_j \Big)} - C_0 \\ 
                 &=\!\sum_{j=1}^{k}{\!\Big(\! \sum_{i=0}^{m-1}{(W_j\, \text{mod}\,2^{m-i})\!\cdot\! a_i\!\cdot\! 2^i}} \!-\! x_j\! \cdot\! \mathrm{E}[\widehat{W}_j] \Big)\! -\! C_0.
\end{aligned}   
\label{eq:axconverr_tr}\raisetag{45pt}
\end{gather}
Then, the mean convolution error $\mathrm{E}[\epsilon_{G^\ast}]$ is given by:
\begin{equation}\label{eq:axconvmean_tr}
\begin{split}
\mathrm{E}[\epsilon_{G^\ast}] &= \sum_{j=1}^{k}{\mathrm{E}\Big[\epsilon_j - v_j \Big]} - C_0\\
                     &= \sum_{j=1}^{k}{\Big( \sum_{i=0}^{m-1}{(W_j \text{ mod }2^{m-i})\cdot 2^i \cdot \mathrm{E}[a_i]}}\Big)\\
                     & \quad - \sum_{j=1}^{k}{\mathrm{E}[x_j]\cdot \mathrm{E}[\widehat{W}_j]} - C_0\\
                     & = \sum_{j=1}^{k}{\widehat{W}_j} - \underbrace{\frac{2^m-1}{2^m}}_{\mathrm{E}[x_j]}\sum_{j=1}^{k}{\widehat{W}_j} - C_0\\
                     & = \frac{1}{2^m}\sum_{j=1}^{k}{\widehat{W}_j} - C_0  
\end{split}
\end{equation}
Therefore, by setting $C_0=\frac{1}{2^m}\sum_{j=1}^{k}{\widehat{W}_j}$,~\eqref{eq:axconvmean_tr} becomes zero.
As a result, in the case of the approximate truncated multiplier, the proposed control variate approximation with $V= \sum_{j=1}^{k}{\mathrm{E}[\widehat{W}_j](1-\delta_{0,y_j})}+\frac{1}{2^m}\sum_{j=1}^{k}{\widehat{W}_j}$ nullifies the mean error.
Considering also that the error variance of the truncated multiplier is small, the convolution error variance will also be limited.
Note that in this case $C_0\neq0$.
However, the addition of $C_0$ is performed with zero cost by just updating offline the bias value of the respective filter as in~\cite{tasoulas2020weight}.

\subsection{Control Variate with Approximate Recursive Multipliers}\label{subsec:covarrec}
Finally, we present the respective control variate analysis when the approximate recursive multipliers are employed, i.e., $\mathrm{AM}(W_j,A_j) = \mathrm{AM_R}(W_j,A_j)$.
Considering the error of an approximate perforated multiplication that is given by~\eqref{eq:err_rec}, we set $x_j$ in our control variate~\eqref{eq:v2} equal to:
\begin{equation}\label{eq:v_rec}
x_j = A_j\text{ mod }2^m = A_j\, \&\, (2^m-1)
\end{equation}
and thus, in the case of the approximate recursive multipliers~\eqref{eq:axconverr} becomes:
\begin{equation}\label{eq:axconverr_rec}
\begin{split}
\epsilon_{G^\ast}&=\sum_{j=1}^{k}{\Big(\epsilon_j - v_j \Big)} - C_0 \\
                 &=\sum_{j=1}^{k}{\Big(x_j\cdot((W_j \text{ mod } 2^m) -C) \Big)} - C_0.
\end{split}            
\end{equation} 
By denoting
\begin{equation}
W_j^m=W_j \text{ mod } 2^m,
\end{equation}
we set $C=\mathrm{E}[W_j^m]\cdot x_j$ and $C_0=0$  in our control variate:
\begin{equation}\label{eq:v2_rec}
V=\sum_{j=1}^{k}{\mathrm{E}[W_j^m]\cdot x_j}, \quad x_j = A_j\text{ mod }2^m.
\end{equation}
Hence, the mean error of the approximate convolution ($\mathrm{E}(\epsilon_{G^\ast})$) is nullified and the variance of the approximate convolution error ($\mathrm{Var}(\epsilon_{G^\ast})$) is minimized.
Proofs are identical to the perforated multiplier case and thus omitted.

\subsection{Application with Other Approximate Multipliers}
As our analysis in Sections~\ref{subsec:covarperf} to \ref{subsec:covarrec} demonstrates, our control variate technique can be effectively applied with three diverse approximate multipliers that exhibit varying error characteristics.
The examined approximate multipliers apply high approximation leading to high power savings but also high error.
Still, our rigorous error analysis shows that our method is able to mitigate the induced error at the convolution level.
Our proposed control variate approximation is not limited to just the examined multipliers, but it can be employed with any approximate multiplier as long as its error can be expressed by an analytical model.
Nevertheless, the cost-efficiency of our approach will depend on the complexity associated with computing the error model.
However, as our analysis for the truncated multiplier demonstrates, our approach can still be efficiently applied, even in the case of a complex error model, as long as i) we can assess with low-cost if a multiplication error occurred and ii) the error of the approximate multiplier features low variance.

\section{Approximate DNN Accelerator}\label{sec:architecture}

\begin{figure}[t!]
\centering
\resizebox{0.95\columnwidth}{!}{\includegraphics{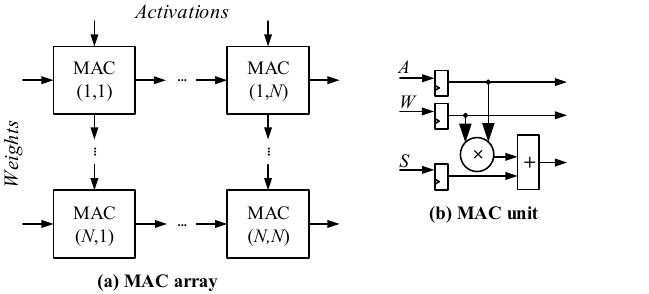}}
\caption{The a) accurate systolic MAC array and b) MAC unit.
Figure obtained from~\cite{zervakis2021control}.}
\label{fig:acmacarray}
\end{figure}

\begin{figure}[t!]
\centering
\resizebox{0.95\columnwidth}{!}{\includegraphics{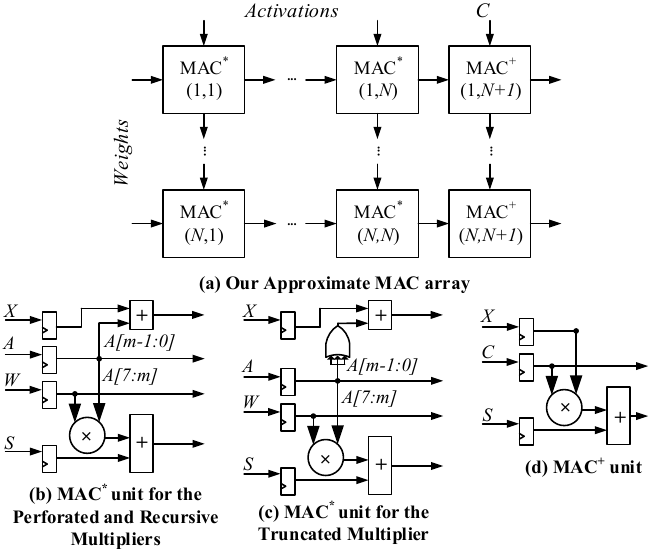}}
\caption{a) Our approximate systolic MAC array. b) MAC$^*$ unit for the Approximate Perforated and Recursive Multipliers. c) MAC$^*$ unit for the Approximate Truncated Multiplier. d) MAC$^+$ unit.
Figure modified from~\cite{zervakis2021control}.}
\label{fig:axmacarray}
\end{figure}

We employ our proposed control variate technique and design approximate DNN accelerators based on a micro-architecture similar to the Google TPU~\cite{jouppi2017datacenter}.
The latter is composed of a large $N\times N$ systolic MAC array, as the one depicted in Fig.~\ref{fig:acmacarray}a.
Fig.~\ref{fig:acmacarray}b presents a pipelined accurate MAC unit, i.e., the processing element replicated within the array.
Each MAC unit comprises an $8$-bit multiplier and a $\lceil \log_2( N \times (2^{16}-1))\rceil$-bit adder to avoid accumulation overflow~\cite{tasoulas2020weight}.
As an example, for a $64\times 64$ MAC array, the size of the adder is $22$-bit.
In the approximate MAC array, depicted in Fig.~\ref{fig:axmacarray}a, each accurate MAC unit is substituted by its approximate version MAC$^*$, where the accurate multiplier is replaced with an approximate one from Section~\ref{sec:axmults}.
Moreover, the MAC$^*$ unit is enhanced with the circuit computing the partial sum of $\sum_{j=1}^{k}{x_j}$ needed to calculate the control variate $V$ (see~\eqref{eq:v2}) of each row.
Finally, the approximate systolic array needs $N+1$ columns, i.e. one more column with respect to the accurate design.
The extra column is composed by MAC$^+$ units.
The MAC$^+$ unit calculates $V=C\cdot\sum_{j=1}^{k}{x_j}$ and adds it to the convolution result generated by the MAC$^*$ in the first $N$ columns: $B+\sum_{j=1}^{k}{\mathrm{AM}(W_j,A_j)}$.
As discussed in the previous Section, the control variate approximation depends on the selected approximate multiplier, so the hardware implementation of the MAC$^*$ unit is modified accordingly. 

\subsection{MAC$^*$ unit with Approximate Perforated Multipliers}\label{subsec:macs_perf}
The product of the approximate perforated multiplier requires $16-m$ bits since $m$ partial products are omitted.
Hence, the adder of MAC$^*$ can be simplified since its size can be reduced by $m$ bits accordingly (compared with the adder of the MAC unit).
The adder within the MAC$^*$ units of the first column adds the first $8-m$ MSBs of $B$, i.e. $B[7\!:\!m]$, to the first approximate product $\mathrm{AM_P}(W_1,A_1)$.
The addition of $B[m\!-\!1\!:\!0]$ occurs in the MAC$^+$ unit, as it will be explained later.
Moreover, each MAC$^*$ needs an extra adder to compute the partial sum required to calculate $V$.
As explained in Section~\ref{subsec:covarperf}, for the perforate multiplier $x_j=A_{j}[m\!-\!1\!:\!0]$ and is $m-$bit wide.
Thus, a $\lceil \log_2( N \times (2^m-1))\rceil$-bit adder is required.
It is worth noting that such a size is considerably lower than the size of the main adder.
As an example, for $N=64$ and $m=2$, the size of the extra adder in the MAC$^*$ unit is $8$ bits while the main adder in MAC is $22$ bits wide.
Therefore, the associated hardware overhead is very small.          

Overall, each MAC$^*$ belonging to the $h$-th column of the approximate MAC array computes the following:
\begin{equation}\label{eq:macstar_perf}
\begin{gathered}
    P_h^*=\mathrm{AM_P}(W_h,A_h)\\
    sum_h=sum_{h-1}+P_h^*\,,\ sum_0=B[7:m]\\
    sumX_h=sumX_{h-1}+A_h[m-1:0]\,,\ sumX_0=0
\end{gathered}
\end{equation}

\subsection{MAC$^*$ unit with Approximate Truncated Multipliers}
The output of the $8\times 8$ approximate truncated multiplier requires $16-m$ bits, since the $m$ least significant columns are truncated.
Consequently, as for the approximate perforated MAC$^*$, the main adder in the MAC$^*$ unit can be again reduced by $m$ bits with respect to the main adder of the MAC.
Similarly, the main adder of the MAC$^*$ units, belonging to the first column, receives, as one of the two inputs, the number $B[7\!:\! m]$.
As described in Section~\ref{subsec:covartrunc}, in the case of the approximate truncated multiplier $x_j$ is $1$-bit wide and it is equal to $(1-\delta_{0,y_j})$, where $y_j = A_j\,\text{mod}\,2^m$.
Hence, to calculate $x_j$, a simple $m-$bit \texttt{OR} gate is required that receives the bits $A_j[m\!-\!1\!:\!0]$ as inputs.
The output of the \texttt{OR} gate is sent to a small adder that calculates the partial sum of $\sum_{j=1}^{k}{x_j}$.
Since a MAC$^*$ can increment the latter by at most $1$, the size of the small adder does not depend on $m$ and it is equal to $\lceil \log_2( N )\rceil$.
Overall, each MAC$^*$ belonging to the $h$-th column of the approximate MAC array computes the following:

\begin{gather}
\begin{gathered}
    P_h^*=\mathrm{AM_T}(W_h,A_h)\\
    sum_h=sum_{h-1}+P_h^*\,,\ sum_0=B[7:m]\\
    sumX_h=sumX_{h-1}+\texttt{OR}(A_h[m\!-\!1\!:\!0]),\ sumX_0=0
\end{gathered} \label{eq:macstar_trunc}\raisetag{27pt}
\end{gather}

\subsection{MAC$^*$ unit with Approximate Recursive Multipliers}
In the approximate recursive multiplier the sub-product $W_L\times A_L$ is omitted, with $W_L$ and $A_L$ being $m-$bit wide.
Hence, again the approximate product requires $16\!-\!m$ bits.
As described in~\ref{subsec:covarrec}, $x_j=A_j[m\!-\!1\!:\!0]$, i.e., exactly the same in the MAC$^*$ of the approximate perforated multiplier.
Therefore, the design of the MAC$^*$ unit of the approximate recursive multiplier is similar to Section~\ref{subsec:macs_perf} and computes:
\begin{equation}\label{eq:macstar}
\begin{gathered}
    P_h^*=\mathrm{AM_R}(W_h,A_h)\\
    sum_h=sum_{h-1}+P_h^*\,,\ sum_0=B[7:m]\\
    sumX_h=sumX_{h-1}+A_h[m-1:0]\,,\ sumX_0=0
\end{gathered}
\end{equation}

\subsection{MAC$^+$ unit}
Each MAC$^+$ unit (last column of the systolic array) calculates the control variate $V=C\cdot\sum_{j=1}^{N}{x_j}$ by means of an exact multiplier.
This is a common feature regardless the approximate multiplier employed in the MAC$^*$ unit.
The only difference is the size of the multiplier: it is equal to $\lceil \log_2( N \times (2^m-1))\rceil\times 8$ when the approximate perforated and recursive multipliers are used in MAC$^*$, while it is equal to $\lceil \log_2( N)\rceil\times 8$ for the truncated one.
Furthermore, a $\lceil \log_2( N \times (2^{16}-1))\rceil$-bit adder is required to produce the final output $G^\ast$. Basically, the MAC$^+$ unit calculates the following:
\begin{gather}
    V=C\cdot sumX_{N}  \label{eq:macplusP} \\
    G^\ast=\{sum_N,B[m-1:0]\}+V \label{eq:macplusS} 
\end{gather}
It is noteworthy that by concatenating $sum_N$ and $B[m\!-\!1\!:\!0]$ in~\eqref{eq:macplusS} we manage to: i) shift left $m$ places the partial sum of the MAC$^*$ units ($sum_N$) as required, and ii) add the $m$-LSBs of the bias $B$ ($B[m\!-\!1\!:\!0]$), which were not taken into account in the MAC$^*$.
The additional $(N+1)-$th column composed of MAC$^+$ units increases the latency of the MAC array by one cycle.
If the delay of the MAC$^+$ is higher than the delay of the accurate MAC, the MAC$^+$ unit can be further pipelined in order to sustain the same operating frequency.
In this case, the latency may increase by two clock cycles per convolution layer.
However, this overhead is completely negligible since the inference phase typically requires thousands of cycles for each convolution layer~\cite{amrouch2020npu}.
Finally, the MAC$^+$ requires the value $C$ to calculate $V$.
This value can be transferred to the DNN accelerator along with the weights of the filters.
Considering the size of DNNs, the data transfer overhead is again negligible.

As a final consideration, it is immediate to note that the computations of $sumX_h$ and $sum_h$ are independent so they are executed in parallel.
Therefore, the adder required to calculate $sumX_h$ is not on the critical path of MAC$^*$ and thus, a slower and power-efficient ripple-carry adder can be employed.
Moreover, the application of the selected approximate technique allows reducing the delay of the MAC$^*$ with respect to the exact MAC.
In addition, as explained above pipelining is used to make MAC$^+$ as fast as the exact MAC.
Therefore, this delay slack enables downsizing the gates of the critical paths and boosts further the area and power savings~\cite{VenkataramaniDATE2013}.

\section{Experimental Results}\label{sec:experimental}

In this section, we evaluate the efficiency of our proposed control variate approximation in terms of area, power, and accuracy.
To achieve this, we design several $N\times N$ exact as well as approximate MAC arrays.
Four values are considered for $N$, from $16$ up to $64$.
The accurate and approximate versions of the MAC array have been synthesized using Synopsys Design Compiler and mapped to our $14$nm technology library~\cite{amrouch2020npu}.
All the designs are implemented using the optimized components of the Synopsys DesignWare Library (i.e., reduction trees and adders), as typically done in commercial flows.
During the synthesis, the \texttt{compile\_ultra} command has been used to target performance optimization.
The accurate MAC array has been synthesized at its minimum clock period.
The latter value has been also used as a time constraint during the synthesis of the approximate MAC arrays.
As described in the previous section, the MAC$^*$ units are inherently faster than the accurate MAC, so the synthesis of the approximate array has been relaxed allowing optimizing area and power dissipation.
Therefore, in the following evaluation, area and power analysis are performed at iso-delay.
Power consumption is calculated with Synopsys PrimeTime on the basis of post-synthesis back-annotated simulations performed using Mentor Questasim.
We simulate the analyzed MAC arrays for 10,000 inference cycles to obtain precise switching activity estimation.
Such a value of simulated cycles is a good compromise between the accuracy of the obtained power results and the simulation duration time.
Running post-synthesis timing simulations for the entire inference phase is infeasible due to the vast time required~\cite{tasoulas2020weight}. 
The inference accuracy is obtained by integrating the proposed control variate approximation into the approximate TensorFlow implementation of~\cite{tfapproxDate2020}.
For the accuracy evaluation, we consider six popular CNNs of varying size, depth, and architecture, trained on the Cifar-10 and Cifar-100 datasets.

\subsection{Hardware Evaluation}
First, we evaluate the power and area gains of our approximate MAC arrays compared to the accurate ones.
For all the evaluated approximate designs, the critical path of the MAC$^+$ unit is shorter than the critical path of the exact MAC, so it does not need to be further pipelined.
Therefore, compared to the accurate design, our approximate ones exhibit a latency overhead of only one clock cycle per convolution layer.     

\subsubsection{Control Variate \& Approximate Perforated Multipliers}
\begin{figure}[t!]
\centering
\resizebox{\columnwidth}{!}{\includegraphics{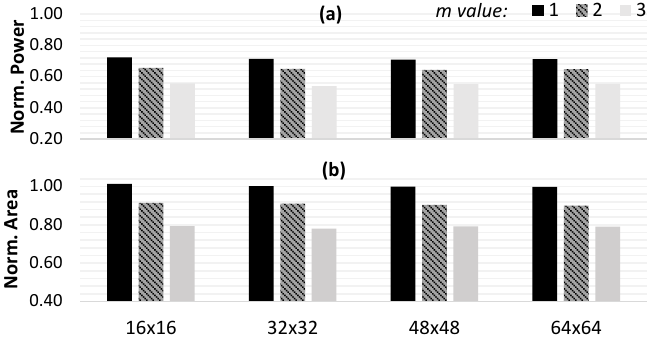}}
\caption{The a) power and b) area of our control variate approximation when using the approximate perforated multipliers for $m\in[1, 3]$ and for varying MAC array sizes. The area and power values are normalized over the corresponding values of the respective accurate design.}
\label{fig:hw_perf}
\end{figure}

Fig.~\ref{fig:hw_perf} presents the area and power savings delivered by the proposed control variate technique when applied with the perforated multiplier for increased values approximation ($m$).
As shown in Fig.~\ref{fig:hw_perf}a, our approximate MAC arrays achieve large power reduction, ranging from $27.7$\% up to $46.1$\%.
As expected, the power gain is directly proportional to the value of $m$.
For $m$=$1$, the power reduction ranges from $27.7$\% to $29.2$\%.
For $m$=$2$, the respective values range from $34.5$\% to $35.7$\%, while for $m$=$3$, the power decreases from $44.4$\% to $46.1$\%.
It is worth noting that the power reduction is almost insensitive to the array dimension $N$. 

Fig.~\ref{fig:hw_perf}b shows that the area reduction entailed by the proposed approximate control variate technique is mainly defined by the $m$ value and is slightly affected by the array size.
As explained in the previous Section, the higher the value of $m$ the higher the approximation and the lower the number of hardware resources, mainly Full Adders (FAs), needed by the MAC$^*$ module.
However, the MAC$^*$ requires more flip-flops (FFs) than the accurate MAC due to the pipeline of the $sumX$ path.
For this reason, the area occupancy of the MAC$^*$ unit for $m$=$1$ is almost the same with the one shown by the accurate MAC.
Contrary, for higher values of $m$ the area saving due to the reduced number of FAs overcomes the drawback of a higher number of FFs.
Indeed, for $m$=$3$, the area gain goes up to $22$\%.
On average, the area reduction is $10$\%.

\begin{figure}[t!]
\centering
\resizebox{\columnwidth}{!}{\includegraphics{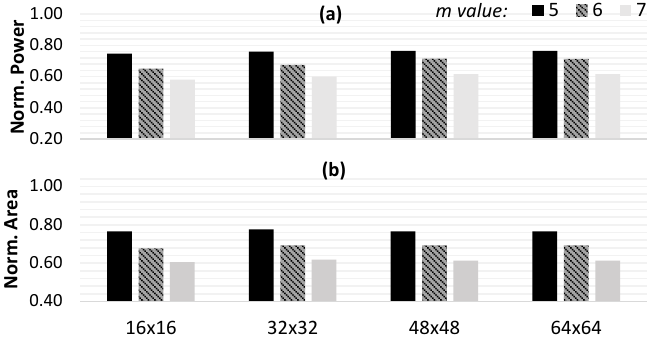}}
\caption{The a) power and b) area of our control variate approximation when using the approximate truncated multipliers for $m\in[5, 7]$ and for varying MAC array sizes. The area and power values are normalized over the corresponding values of the respective accurate design.}
\label{fig:hw_trunc}
\end{figure}
\begin{figure}[t!]
\centering
\resizebox{\columnwidth}{!}{\includegraphics{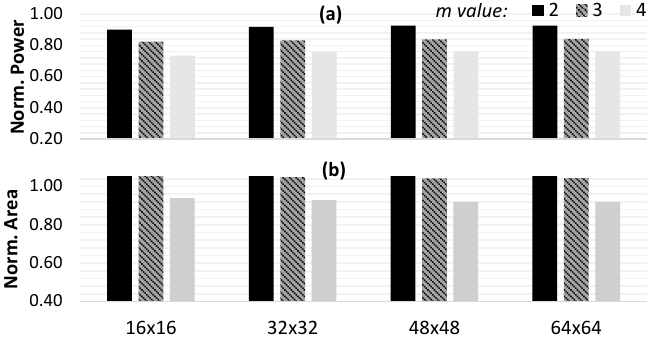}}
\caption{The a) power and b) area of our control variate approximation when using the approximate recursive multipliers for $m\in[2, 4]$ and for varying MAC array sizes. The area and power values are normalized over the corresponding values of the respective accurate design.}
\label{fig:hw_rec}
\end{figure}

\subsubsection{Control Variate \& Approximate Truncated Multipliers}
Fig.~\ref{fig:hw_trunc}a shows the savings attained by our control variate technique when applied with the truncated multiplier.
Similarly, the power dissipation decreases as the number of truncated columns $m$ increases.
For the considered range of $m\in[5, 7]$, the power gain goes up to a maximum of $41.9$\%, obtained for $m$=1 and $N$=16.
In this case, the power sensitivity of $N$ is quite higher when compared to using the perforated multiplier.
For $m$=$1$ the power gain ranges from $23.5$\% to $25.4$\% while for $m$=$2$ the respective values are from $28.6$\% to $35.0$\%.
For $m$=3 the power savings range from $38.4$\% to $41.9$\%.

As shown in Fig.~\ref{fig:hw_trunc}b, the area gain delivered by the control variate technique on the truncated array is considerably high, reaching a maximum value of $39$\% for $m$=7. 
On average, the area reduction of the truncated multiplier is $31$\%.
It is worth noting that the truncated multiplier entails a higher area saving with respect to the perforation approximation. 
This is explained by the fact that the additional adder that computes $sumX$ is smaller and the associated FFs are fewer.

\subsubsection{Control Variate \& Approximate Recursive Multipliers}
Finally, the power and area savings when using the approximate recursive multiplier are depicted in Fig.~\ref{fig:hw_rec}a-b.
In this case, the approximate MAC array shows the lowest power gain since the power reduction is up to $26$\% ($17$\% on average) compared to the exact design.
Similarly, the maximum area gain is only $8$\%.
Such limited gains are attributed to the fact that when $m$ is small i) the hardware savings of omitting the least significant sub-product are constrained, and ii) the additional logic required by the control variate is significant compared to the approximation gains.
It is noteworthy that for $m$=$2$ and $N$=$16$ there is a $14$\% area overhead.  

%perforated accuracy
\begin{table}[t!]
\renewcommand{\arraystretch}{1.2}
\caption{Accuracy Evaluation when Considering the Approximate Perforated Multiplier. Six Neural Networks Trained on Cifar-10 and Cifar-100 Datasets Are Examined.
}
\label{tab:accPerforated}
\footnotesize
\centering
\begin{threeparttable}
  \begin{tabular}[t]{l|c|c|c|c|c|c}
  \hline
\multicolumn{7}{c}{\textbf{Accuracy Loss (\%)}} \\ \hline
\multirow{2}{*}{\makecell{\textbf{NN on}\\ \textbf{Cifar-10}}}  & \multicolumn{2}{c|}{$m=1$} & \multicolumn{2}{c|}{$m=2$} & \multicolumn{2}{c}{$m=3$} \\ \cline{2-7}
  & \textbf{Ours}\tnote{+} & \textbf{ w/o $\boldsymbol V$}\tnote{*} & \textbf{Ours} & \textbf{ w/o $\boldsymbol V$} & \textbf{Ours} & \textbf{ w/o $\boldsymbol V$} \\ \hline
GoogLeNet & -0.16 & 0.35 & 0.00 & 4.13 & 1.95 & 31.78 \\ \hline
ResNet44 & 0.03 & 0.73 & 0.83 & 4.49 & 5.75 & 32.94 \\ \hline
ResNet56 & 0.49 & 0.60 & 1.25 & 5.36 & 5.94 & 39.01 \\ \hline
ShuffleNet & 0.12 & 3.40 & -0.48 & 7.60 & 6.53 & 31.74 \\ \hline
VGG13 & -0.01 & 0.23 & -0.30 & 0.76 & 0.69 & 6.76 \\ \hline
VGG16 & -0.13 & 0.19 & 0.38 & 1.66 & 3.86 & 8.71 \\ \hline
\textbf{Average} & 0.06 & 0.92 & 0.28 & 4.00 & 4.12 & 25.16 \\ \hline
\hline
\multirow{2}{*}{\makecell{\textbf{NN on}\\ \textbf{Cifar-100}}}    & \multicolumn{2}{c|}{$m=1$} & \multicolumn{2}{c|}{$m=2$} & \multicolumn{2}{c}{$m=3$} \\ \cline{2-7}
  & \textbf{Ours} & \textbf{ w/o $\boldsymbol V$} & \textbf{Ours} & \textbf{ w/o $\boldsymbol V$} & \textbf{Ours} & \textbf{ w/o $\boldsymbol V$} \\ \hline
GoogLeNet & 0.05 & 2.43 & 1.47 & 13.19 & 7.02 & 44.52 \\ \hline
ResNet44 & 0.77 & 1.02 & 1.82 & 14.17 & 11.27 & 43.40 \\ \hline
ResNet56 & -0.34 & 3.02 & 2.25 & 15.83 & 14.34 & 44.59 \\ \hline
ShuffleNet & 0.20 & 9.47 & 1.09 & 5.92 & 6.57 & 15.84 \\ \hline
VGG13 & 0.89 & 3.01 & 1.91 & 5.89 & 5.52 & 14.70 \\ \hline
VGG16 & -0.03 & 3.96 & 0.03 & 2.41 & 2.80 & 11.54 \\ \hline
\textbf{Average} & 0.26 & 3.82 & 1.43 & 9.57 & 7.92 & 29.10 \\ \hline
  \end{tabular}
\begin{tablenotes}\footnotesize
\item[+] Accuracy achieved when using the Approximate Perforated Multiplier with our proposed control-variate approximation.
\item[*] Accuracy when using only the Approximate Perforated Multiplier.
\end{tablenotes}
\end{threeparttable}
\end{table}

%truncated accuracy
\begin{table}[t!]
\renewcommand{\arraystretch}{1.2}
\caption{Accuracy Evaluation when Considering the Approximate Truncated Multiplier. Six Neural Networks Trained on Cifar-10 and Cifar-100 Datasets Are Examined.
}
\label{tab:accTruncated}
\footnotesize
\centering
\begin{threeparttable}
  \begin{tabular}[t]{l|c|c|c|c|c|c}
  \hline
\multicolumn{7}{c}{\textbf{Accuracy Loss (\%)}} \\ \hline
\multirow{2}{*}{\makecell{\textbf{NN on}\\ \textbf{Cifar-10}}}  & \multicolumn{2}{c|}{$m=5$} & \multicolumn{2}{c|}{$m=6$} & \multicolumn{2}{c}{$m=7$} \\ \cline{2-7}
  & \textbf{Ours}\tnote{+} & \textbf{ w/o $\boldsymbol V$}\tnote{*} & \textbf{Ours} & \textbf{ w/o $\boldsymbol V$} & \textbf{Ours} & \textbf{ w/o $\boldsymbol V$} \\ \hline
GoogLeNet & -0.44 & 1.54 & 0.19 & 10.38 & 0.79 & 33.78 \\ \hline
ResNet44 & 0.20 & 0.98 & 0.48 & 4.95 & 3.93 & 27.70 \\ \hline
ResNet56 & -0.37 & 0.76 & 0.22 & 5.23 & 5.00 & 35.07 \\ \hline
ShuffleNet & 1.13 & 3.51 & 1.89 & 14.26 & 16.67 & 49.55 \\ \hline
VGG13 & 0.07 & 5.14 & 4.14 & 35.19 & 17.33 & 72.42 \\ \hline
VGG16 & 1.21 & 5.65 & 13.84 & 41.42 & 33.98 & 74.95 \\ \hline
\textbf{Average} & 0.30 & 2.93 & 3.46 & 18.57 & 12.95 & 48.91 \\ \hline
\hline
\multirow{2}{*}{\makecell{\textbf{NN on}\\ \textbf{Cifar-100}}}    & \multicolumn{2}{c|}{$m=5$} & \multicolumn{2}{c|}{$m=6$} & \multicolumn{2}{c}{$m=7$} \\ \cline{2-7}
  & \textbf{Ours} & \textbf{ w/o $\boldsymbol V$} & \textbf{Ours} & \textbf{ w/o $\boldsymbol V$} & \textbf{Ours} & \textbf{ w/o $\boldsymbol V$} \\ \hline
GoogLeNet & 0.58 & 11.84 & 1.13 & 24.62 & 9.45 & 53.04 \\ \hline
ResNet44 & -0.60 & 3.39 & 0.67 & 11.25 & 9.29 & 36.66 \\ \hline
ResNet56 & -0.70 & 2.50 & 1.00 & 13.96 & 8.84 & 40.29 \\ \hline
ShuffleNet & -2.63 & 10.43 & -0.85 & 23.41 & 8.22 & 37.50 \\ \hline
VGG13 & 3.18 & 18.83 & 5.68 & 53.85 & 27.78 & 64.70 \\ \hline
VGG16 & 0.70 & 20.31 & 6.96 & 56.32 & 36.81 & 63.52 \\ \hline
\textbf{Average} & 0.09 & 11.22 & 2.43 & 30.57 & 16.73 & 49.29 \\ \hline
  \end{tabular}
\begin{tablenotes}\footnotesize
\item[+] Accuracy achieved when using the Approximate Truncated Multiplier with our proposed control-variate approximation.
\item[*] Accuracy when using only the Approximate Truncated Multiplier.
\end{tablenotes}
\end{threeparttable}
\end{table}

%recursive accuracy
\begin{table}[t!]
\renewcommand{\arraystretch}{1.2}
\caption{Accuracy Evaluation when Considering the Approximate Recursive Multiplier. Six Neural Networks Trained on Cifar-10 and Cifar-100 Datasets Are Examined.
}
\label{tab:accRecursive}
\footnotesize
\centering
\begin{threeparttable}
  \begin{tabular}[t]{l|c|c|c|c|c|c}
  \hline
\multicolumn{7}{c}{\textbf{Accuracy Loss (\%)}} \\ \hline
\multirow{2}{*}{\makecell{\textbf{NN on}\\ \textbf{Cifar-10}}}  & \multicolumn{2}{c|}{$m=2$} & \multicolumn{2}{c|}{$m=3$} & \multicolumn{2}{c}{$m=4$} \\ \cline{2-7}
  & \textbf{Ours}\tnote{+} & \textbf{ w/o $\boldsymbol V$}\tnote{*} & \textbf{Ours} & \textbf{ w/o $\boldsymbol V$} & \textbf{Ours} & \textbf{ w/o $\boldsymbol V$} \\ \hline
GoogLeNet & -0.11 & 0.15 & 0.12 & 1.04 & 0.13 & 3.17 \\ \hline
ResNet44 & -0.10 & -0.03 & 0.02 & 0.07 & 0.11 & 3.82 \\ \hline
ResNet56 & -0.12 & 0.74 & 0.07 & 0.13 & -0.22 & 2.93 \\ \hline
ShuffleNet & -0.28 & -0.28 & 0.07 & 2.70 & 1.59 & 9.77 \\ \hline
VGG13 & -0.48 & 0.01 & -0.81 & 0.54 & 2.18 & 8.11 \\ \hline
VGG16 & 0.09 & 0.41 & 0.31 & 1.89 & 3.13 & 7.85 \\ \hline
\textbf{Average} & -0.17 & 0.17 & -0.04 & 1.06 & 1.15 & 5.94 \\ \hline
\hline
\multirow{2}{*}{\makecell{\textbf{NN on}\\ \textbf{Cifar-100}}}    & \multicolumn{2}{c|}{$m=2$} & \multicolumn{2}{c|}{$m=3$} & \multicolumn{2}{c}{$m=4$} \\ \cline{2-7}
  & \textbf{Ours} & \textbf{ w/o $\boldsymbol V$} & \textbf{Ours} & \textbf{ w/o $\boldsymbol V$} & \textbf{Ours} & \textbf{ w/o $\boldsymbol V$} \\ \hline
GoogLeNet & -0.09 & -0.09 & 1.58 & 2.00 & 0.53 & 19.19 \\ \hline
ResNet44 & 0.13 & 0.13 & -1.00 & 1.87 & 0.03 & 9.58 \\ \hline
ResNet56 & -0.32 & -0.32 & -0.30 & 0.16 & 0.25 & 10.63 \\ \hline
ShuffleNet & -4.08 & -3.49 & 1.17 & -1.10 & -0.12 & 20.50 \\ \hline
VGG13 & 2.33 & 8.75 & 4.18 & 8.43 & 9.26 & 38.65 \\ \hline
VGG16 & -0.59 & -0.59 & -2.65 & 3.67 & 8.65 & 35.74 \\ \hline
\textbf{Average} & -0.44 & 0.73 & 0.50 & 2.51 & 3.10 & 22.38 \\ \hline
  \end{tabular}
\begin{tablenotes}\footnotesize
\item[+] Accuracy achieved when using the Approximate Recursive Multiplier with our proposed control-variate approximation.
\item[*] Accuracy when using only the Approximate Recursive Multiplier.
\end{tablenotes}
\end{threeparttable}
\end{table}

\subsection{Accuracy Evaluation}
In this Section we evaluate the impact of our control variate technique on the delivered inference accuracy.
Tables~\ref{tab:accPerforated}-\ref{tab:accRecursive} report the accuracy loss of our control variate approximation compared to the exact design, for the examined approximate multipliers and for varying approximation values ($m$), over six CNNs.
Note that the accuracy does not depend on the size of the MAC array, since $N$ only affects the number of operations performed in parallel by the array.
Moreover, Tables~\ref{tab:accPerforated}-\ref{tab:accRecursive} also report the accuracy loss when the approximate multiplier is used in the inference without our control variate technique (i.e. without adding $V$).
The latter highlights the accuracy improvement that is delivered by our method.
Negative values in Tables~\ref{tab:accPerforated}-\ref{tab:accRecursive} refer to higher accuracy compared to the baseline~\cite{ansari2019improving,alwann}.

As shown in Table~\ref{tab:accPerforated}, the average accuracy loss of our method when using the approximate perforated multiplier for the Cifar-10 dataset is $0.06$\%, $0.28$\%, and $4.12$\% for $m=1$, $m=2$, and $m=3$ respectively.
The corresponding values for the more challenging Cifar-100 dataset are $0.26$\%, $1.43$\%, and $7.92$\%.
As a result, our technique achieves \mytilde$24$\% power reduction for negligible accuracy loss, i.e., $0.16$\% on average on both datasets for $m=1$.
The power gains rise to \mytilde$36$\% ($m=2$) for an average accuracy loss of only $0.85$\%.
Finally, for $6.02$\% average accuracy loss ($m=3$), the power savings jump to \mytilde$46$\%.
In addition, Table~\ref{tab:accPerforated} highlights the efficiency of our control variate approximation in decreasing the convolution error.
Indeed, compared with using the perforated multiplier standalone (i.e., without adding $V$), our technique achieves $2$\%, $6$\%, and $21$\% higher accuracy, on average, for $m=1$, $m=2$, and $m=3$ respectively.
Interestingly, the higher the value of $m$ the higher the accuracy improvement.
This proves that our proposed control variate approach is also very effective when the approximation configuration of the employed approximate multiplier causes a large error value.
As a consequence, the control variate technique can enable an aggressive approximation, and thus a significant energy reduction at a reduced accuracy loss.  

\begin{figure*}[t]
\centering
\includegraphics{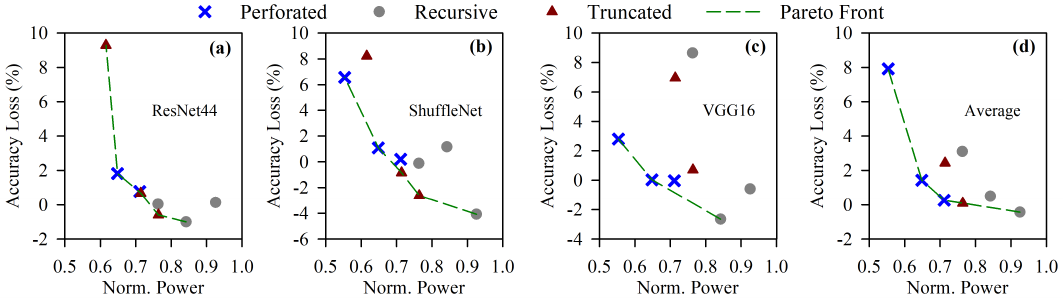}
\caption{The Accuracy Loss (\%) -- Normalized Power Pareto space for varying CNNs (subfigures a-c) as well as the average case (subfigure d). 
Cifar-100 and $N$=$64$ are used.
Normalized Power and Accuracy Loss are reported w.r.t. the baseline exact design.}
\label{fig:PARETO}
\end{figure*}

Similarly, Table~\ref{tab:accTruncated} presents the accuracy evaluation when using the approximate truncated multiplier.
The average accuracy loss of our method for the Cifar-10 dataset is $0.3$\%, $3.46$\%, and $12.95$\% for $m=5$, $m=6$, and $m=7$ respectively.
The corresponding values for the Cifar-100 dataset are $0.09$\%, $2.43$\%, and $16.73$\%.
Moreover, compared to the case of using the approximate truncated multiplier standalone (i.e., without adding $V$), the proposed control variate technique improves the obtained accuracy by up to $22$x for the same $m$ value.
On average, over all the examined cases, our control variate method improves the accuracy by $3.32$x.
As a result, the control variate technique applied to the truncated multiplier can reduce the power dissipation by \mytilde$25$\% at a negligible accuracy loss of only $0.2$\% on average for $m$=$5$.

Finally, the same trend is observed in Table~\ref{tab:accRecursive} for the approximate recursive multiplier.
Similarly to the previous cases, the accuracy loss is proportional to the value of $m$.
Our control variate approach applied to the approximate recursive multipliers entails the lowest accuracy loss among the analyzed approximate multipliers.
Indeed, the maximum accuracy loss is only $3.1$\% on average for $m$=4.
The average accuracy loss for the Cifar-10 dataset is $-0.17$\%, $-0.04$\%, and $1.15$\% for $m=2$, $m=3$, and $m=4$, respectively.
The corresponding values for the Cifar-100 dataset are $-0.44$\%, $0.5$\%, and $3.1$\%.
Compared to using the approximate recursive multiplier standalone (i.e., without adding $V$), our method achieves a maximum accuracy improvement of $2.1$x, obtained for the Cifar-100 dataset and $m$=4.
On average, over all the examined cases, our control variate method improves the accuracy by $4.78$\%.
Finally, the power saving is \mytilde$17$\% at a negligible average accuracy loss of $0.23$\% ($m$=3).
For $m$=4, the power saving is \mytilde$25$\% at the cost of an accuracy loss of \mytilde$2.1$\%.

As it can be inferred from the above discussion, our control variate method always improves the obtained accuracy and enables achieving high power reduction even under more constrained accuracy loss thresholds.
Nevertheless, the power-accuracy trade-off depends on several parameters such as the approximate multiplier type, the applied level of approximation ($m$), and the examined network.
As an example, as shown in Table~\ref{tab:accPerforated} and Table~\ref{tab:accTruncated}, our control variate method on the perforated multiplier mainly delivers, on average, better accuracy than when it is applied with the truncated one.
Though, if we restrict our analysis on the ResNet family (ResNet44 and ResNet56), using the truncated multiplier outperforms the perforated one in most of the cases, for both the Cifar-10 and the Cifar-100.
Fig.~\ref{fig:PARETO} depicts the respective Pareto space (accuracy vs. power) for a variety of representative cases: the a) ResNet44, b) ShuffleNet, c) VGG16, and d) average accuracy over all the examined CNNs.
For the analysis in Fig.~\ref{fig:PARETO}, we consider the Cifar-100 dataset and a $64\times 64$  MAC array size.
In addition, only the configurations that result in up to $10$\% accuracy loss are depicted.
Similar results are obtained for Cifar-10 and the rest of the $N$ values.
As Fig.~\ref{fig:PARETO} shows, there is not a dominant solution, but the optimal design choice depends on the target accuracy threshold and the network's topology.
The Pareto front spans across a wide range of approximate multipliers and configurations.
As a general observation, it is better to apply our control variate approach with the recursive approximate multiplier if the accuracy loss constraint is very tight.
Under relaxed accuracy constraints, the perforated multiplier should be preferred since it delivers mainly the highest power savings.
The truncated multiplier is usually in the middle delivering 
a considerable power reduction for limited accuracy loss.
However, in the case of VGG16, there is not any Pareto-optimal approximate design that uses the truncated multiplier.
Finally, it is important to emphasize that our control variate technique is applied to different approximate multipliers allowing to obtain design points in the Pareto front that would be impossible to reach with just a single approximate multiplier.
Hence, the versatility of our approach leads to more fine-grained traversal of the accuracy-power design space. 

\begin{table*}[t!]
\renewcommand{\arraystretch}{1.2}
\setlength{\tabcolsep}{4pt}
\caption{Evaluation of the Area and Power Overheads of MAC$^+$}
\label{tab:macplus}
  \footnotesize
  
\begin{minipage}{.3\textwidth}
\centering
  \begin{tabular}[t]{c|c|c|c|c}
 \multicolumn{5}{c}{\textit{Approximate Perforated Multiplier in MAC$^*$}} \\ \hline
  \multicolumn{5}{c}{\textbf{Percentage of Total Area (\%)}} \\ \hline
$m$ & $16\times 16$ & $32\times 32$ & $48\times 48$ & $64\times 64$ \\ \hline
1 & 1.07 & 0.55 & 0.38 & 0.28\\ \hline
2 & 1.18 & 0.61 & 0.41 & 0.31\\ \hline
3 & 1.36 & 0.71 & 0.47 & 0.36\\ \hline
  \multicolumn{5}{c}{\textbf{Percentage of Total Power (\%)}} \\ \hline
$m$ & $16\times 16$ & $32\times 32$ & $48\times 48$ & $64\times 64$ \\ \hline
1 & 1.22 & 0.63 & 0.43 & 0.32\\ \hline
2 & 1.32 & 0.68 & 0.46 & 0.35\\ \hline
3 & 1.52 & 0.80 & 0.53 & 0.40\\ \hline
  \end{tabular}
\end{minipage}%
\hspace{0.03\textwidth}
\begin{minipage}{.3\textwidth}
  \footnotesize
\centering
  \begin{tabular}[t]{c|c|c|c|c}
 \multicolumn{5}{c}{\textit{Approximate Recursive Multiplier in MAC$^*$}} \\ \hline
  \multicolumn{5}{c}{\textbf{Percentage of Total Area (\%)}} \\ \hline
$m$ & $16\times 16$ & $32\times 32$ & $48\times 48$ & $64\times 64$ \\ \hline
2 & 0.95 & 0.48 & 0.33 & 0.25 \\ \hline
3 & 1.02 & 0.53 & 0.36 & 0.27 \\ \hline
4 & 1.15 & 0.59 & 0.41 & 0.31 \\ \hline
  \multicolumn{5}{c}{\textbf{Percentage of Total Power (\%)}} \\ \hline
$m$ & $16\times 16$ & $32\times 32$ & $48\times 48$ & $64\times 64$ \\ \hline
2 & 0.99 & 0.50 & 0.34 & 0.25 \\ \hline
3 & 1.06 & 0.54 & 0.36 & 0.27 \\ \hline
4 & 1.16 & 0.58 & 0.39 & 0.30 \\ \hline
  \end{tabular}
\end{minipage}%
\hspace{0.03\textwidth}
\begin{minipage}{.3\textwidth}
  \footnotesize
\centering
  \begin{tabular}[t]{c|c|c|c|c}
 \multicolumn{5}{c}{\textit{Approximate Truncated Multiplier in MAC$^*$}} \\ \hline
  \multicolumn{5}{c}{\textbf{Percentage of Total Area (\%)}} \\ \hline
$m$ & $16\times 16$ & $32\times 32$ & $48\times 48$ & $64\times 64$ \\ \hline
5 & 1.15 & 0.59 & 0.40 & 0.30 \\ \hline
6 & 1.30 & 0.66 & 0.45 & 0.33 \\ \hline
7 & 1.38 & 0.71 & 0.48 & 0.36 \\ \hline
  \multicolumn{5}{c}{\textbf{Percentage of Total Power (\%)}} \\ \hline
$m$ & $16\times 16$ & $32\times 32$ & $48\times 48$ & $64\times 64$ \\ \hline
5 & 1.06 & 0.54 & 0.37 & 0.28 \\ \hline
6 & 1.19 & 0.60 & 0.39 & 0.29 \\ \hline
7 & 1.25 & 0.64 & 0.43 & 0.32 \\ \hline
  \end{tabular}
\end{minipage}
\end{table*}

\subsection{MAC$^+$ Overhead Discussion}
As described in Section~\ref{sec:architecture}, the proposed control variate technique requires an additional column of MAC$^+$ units in order to add $V$ to the final accumulated value, and thus mitigate the accuracy loss due to the approximate multiplications.
Although in the analysis in Fig.~\ref{fig:hw_perf}-\ref{fig:hw_rec} we have evaluated the area and power consumption of the entire approximate MAC array, i.e., including the MAC$^+$ units, it is useful to disaggregate the area and power consumption of the extra MAC$^+$ modules from the total area and power consumption, in order to analyze their impact on the overall architecture.
Table~\ref{tab:macplus} shows the area and power of the MAC$^+$ units expressed as percentages over the total area and power consumption of the approximate systolic MAC array.
As shown in Table~\ref{tab:macplus}, the impact of the MAC$^+$ units is negligible.
The maximum area (power) overhead has been found to be just $1.38$\% ($1.52$\%) for the truncated (perforated) multiplier and array size of $16\times 16$.
As expected, the area overhead increases as the applied approximation ($m$) increases.
This is mainly explained by the fact that as $m$ increases, the area gains in the MAC$^*$ units increase significantly while the MAC$^+$ are hardly affected.
Hence, considering the significantly higher number of MAC$^*$, the area overhead increases with $m$, being however $1.52$\% at maximum.
Moreover, the area overhead decreases as $N$ increases.
This is mainly explained by the fact that the total area of the MAC$^+$ units scales linearly with respect to $N$ while the total area scales quadratically with $N$.
The same conclusion can be drawn about the impact of the extra MAC$^+$ modules on the power consumption of the systolic array, and the its dependency on $m$ and $N$.      
Concluding, Fig.~\ref{fig:hw_perf}, Fig.~\ref{fig:hw_trunc}, and Table~\ref{tab:macplus} clearly demonstrate the scalability of our approach.

\section{Related works}

There has been great interest around approximate computing for neural network inference.
\cite{mrazek2016design} employed approximate multipliers to different convolution layers and~\cite{sarwar2018energy} proposed a compact and energy-efficient multiplier-less artificial neuron.
However,~\cite{mrazek2016design,sarwar2018energy} are based on retraining to recover accuracy loss caused by the usage of approximation.
\cite{hanif2019cann} utilized approximate MAC by splitting the addition and limiting the carry propagation.
Nevertheless,~\cite{mrazek2016design,hanif2019cann} are evaluated on the LeNet network, a very shallow architecture which cannot provide the amount of operations recent DNNs do.
Therefore, both of these methods can be deemed inapplicable in modern scenarios which require deeper network architectures. 
The work in~\cite{mrazek2020using} enhances the approximate multipliers library of~\cite{mrazek2017evoapproxsb} and shows that, in simple DNNs, even without retraining, considerable energy gains for a small accuracy loss can be attained.
Still, in more complex DNNs the energy gains are not maintained~\cite{mrazek2020using}.
The authors in~\cite{alwann} propose a non-uniform architecture that utilized approximate multipliers from~\cite{mrazek2017evoapproxsb}.
Their work tunes the weights accordingly and avoids retraining.
However,~\cite{alwann} requires a heterogeneous design and generates a different approximate accelerator per DNN.
Similarly, inspired by big.LITTLE computing, Spantidi et al.~\cite{Spantidi:TETC2023} implemented heterogeneous DNN accelerators that consist of 8-bit NPUs in conjunction with lower bit-width NPUs to enhance overall throughput while reducing energy consumption during NN inference.
In~\cite{Zervakis:TC2022}, the authors use low-precision MAC units with larger NPUs to reduce power density.
In~\cite{zervakis2020design}, approximate multipliers with reconfigurable accuracy at run-time are generated.
Similar to~\cite{alwann}, they also apply layer-wise approximation, however with limited power reduction.
In~\cite{tasoulas2020weight,spantidi2021pene} approximate reconfigurable multipliers are used, and a mapping strategy is employed to set the approximation level per weight.
\cite{tasoulas2020weight} applies layer-wise approximation while~\cite{spantidi2021pene} introduced a more fine-grained filter-wise approximation.
However,~\cite{tasoulas2020weight,spantidi2021pene} significantly increase the size of the DNN to store the required configurations.
The work in~\cite{riaz2020caxcnn} allows the usage of reduced-complexity multipliers based on Canonic Sign Digit approximation to represent the filter weights of CNNs that have already been trained.
Nevertheless,~\cite{riaz2020caxcnn} requires DNN-specific approximations.
In~\cite{park2021design} the authors present an interleaving method that employs approximate multipliers to minimize energy consumption of MAC-oriented signal processing algorithms with minimal performance loss. Towards improving the multiplication performance for CNN inference,~\cite{hammad9cnn} proposes an architecture comprising a preprocessing precision controller and approximate multiplier designs of varying precision using the static and dynamic segment methods.
Both~\cite{park2021design,hammad9cnn} apply their proposed work on 16-bit inference, while recent CNN accelerators are mostly targeting 8-bit precision~\cite{jouppi2017datacenter}.

\section{Conclusion}
In this work, we introduced control variate approximation to increase the accuracy of approximate DNN accelerators without requiring any DNN retraining.
Our mathematical analysis demonstrates that our technique mitigates the error induced by the approximate multipliers, by effectively nullifying the mean convolution error and reducing its variance. 
As our extensive experimentation over three diverse approximate multipliers, six DNNs, and four MAC array sizes demonstrates, our control variate approximation enables using aggressive approximate multipliers to design approximate DNN accelerators that boost the power savings with limited accuracy loss.

\bibliographystyle{IEEEtran}
%\bibliography{bibliography}

\end{document}